\documentclass[
aps,%
12pt,%
final,%
notitlepage,%
oneside,%
onecolumn,%
nobibnotes,%
nofootinbib,
superscriptaddress,%
noshowpacs,%
centertags]%
{revtex4}
\begin{document}
\selectlanguage{english} 
\title{ON THE NATURE AND THE ORDER OF THE DECONFINING TRANSITION IN QCD.\\
In honour of Yu. A. Simonov on his seventyth birthday.}
\author{Adriano Di Giacomo}
\affiliation{Dipartimento di Fisica dell'Universit\`a di Pisa , and INFN
Sezione di Pisa\\ l.go Pontecorvo 2, I-56127, Pisa, Italy}
%
%
\begin{abstract}

The determination of the parameters of the deconfining transition in
$N_f = 2$  QCD is discussed , and its relevance to the understanding
of the mechanism of color confinement.
\end{abstract}
\maketitle
\section{Introduction.}
Understanding the mechanism by which QCD  confines colored particles is
one of the most fundamental and challenging problems in field
theory\cite{'tH1}\cite{'tH2}.

     Experiments aimed to detect the existence of a deconfining transition at
high temperature by colliding heavy ions have not yet produced a definite
answer\cite{QM02}. Up to now the deconfining transition has been only
observed in numerical simulations of QCD on a lattice\cite{kar1}.
     Identifying a signal of deconfinement is far from trivial in
experiments, and is not easy in lattice simulations either. To be
rigorous confinement means absence of colored particles in asymptotic
states. Of course one cannot check confinement
by looking at all asymptotic states. In the quenched approximation, in
which dynamical quark loops are neglected, one looks at the static
potential acting between a heavy $Q-\bar Q$ pair. The criterion for
confinement is then the existence of a string tension, which means a
linear behaviour at large distances
\begin{equation}
V(r) =\sigma r
\end{equation}

The static potential is measured on the lattice either in terms of the
Wilson loop (parallel transport along a closed path), or equivalently in
terms of the correlator of Polyakov lines (parallel transports along the
time axis). For a Wilson loop of extension  $t$  in the time direction
and  $r$  in space
\begin{equation}
    W(r,t)\mathop\propto_{r,t\to\infty} \exp[-t V(r)]
\end{equation}
Confinement eq(1) implies the so-called area law
\begin{equation}
W(r,t)\propto \exp[-\sigma rt]
\end{equation}
${r,t\to\infty}$ means that they are large compared to the correlation
length.
The correlator of two Polyakov lines can be written,by use of
cluster property,
\begin{equation}
\langle L(r)L^\dagger(0)\rangle\mathop\simeq_{r\to\infty} C\exp[-\sigma r/T]
+|\langle L\rangle|^2
\end{equation}
On the other hand
\begin{equation}
V(r)\simeq -T\ln\langle L(r)L^\dagger(0)\rangle
\end{equation}
It is found by numerical simulations that a critical temperature $T_c$
exists such that for $T<T_c$  $\langle L\rangle$ =0. From eq's(4) and (5)
confinement eq(1) then follows.
    For $T > T_c$ $\langle L\rangle \ne0$ and
\begin{equation}
V(r)\mathop\to_{r,t\to\infty} const
\end{equation}

For quenched $SU(3)$ $T_c/\sqrt(\sigma)\approx.65$ \cite{IW}. A deconfining
phase transition exists at $T_c\simeq270Mev$ and the Polyakov line
$\langle L\rangle $ is the order parameter. The symmetry involved is $Z_3$
which is broken at $T>T_c$, and the transition is order-disorder. Some
problem can arise in the continuum limit with the definition
\cite{KKPZ}of the order  parameter  $\langle L\rangle$ , but the main idea
looks sound.

     The situation is less clear in the more realistic case of full QCD,
including dynamical quarks. There $Z_3$ is not a symmetry anyhow, being
broken by the coupling to the quarks. Moreover the potential at large
distances is not expected to grow with r, due to the conversion of the
potential energy into light $q-\bar q$ pairs (string breaking)\cite{SB}.
     One needs then an alternative criterion for confinement, which ,however,
is not known.

A phase transition is expected anyhow at low quark masses from the low
temperature phase in which chiral symmetry is spontaneously broken, to a
phase in which it is restored: the order parameter for this transition is
the chiral condensate $\langle\bar\psi\psi\rangle$. This transition is
indeed observed on the lattice at $T_c\approx 170Mev$. A priori
restoration of chiral symmetry is not the same as deconfinement, even if
it is physically understandable , e.g. by thinking of a bag model, that
confinement can imply chiral symmetry breaking. In the absence of a
criterion for confinement the question if chiral and deconfining
transition coincide can not even be asked. A third symmetry exists, the
axial U(1) , which is broken at low T as an effect of the anomaly, and is
also expected to be restored at high temperature. In principle the
corresponding transition temperature could be different from that of the
other two transitions.

    An analysis can be done of the chiral transition
based on symmetry arguments and on renormalization group techniques
\cite{PW}. If the assumption is made that the relevant degrees of freedom
at the transition are the pseudoscalar particles an effective lagrangean
can be written on the basis of symmetry and scale invariance, which
describes the density of free energy around the chiral point. For
$N_f\ge3$ no infrared stable fixed point exists , so that the transition
is expected to be first order.
$N_f=2$ is a special case, and the transition can either be weak first
order or second order . If it is first order , it is weak
first order also at $m_q\ne0$; if it is second order the transition at
$m_q\ne0$ is a crossover \cite{PW}. What is the case can be investigated
by numerical simulations on the lattice, even if up to now
\cite{kar3}\cite{ts1}\cite{MILCH} the results have been rather elusive.

     This issue is very important for the understanding of the deconfining
phase transition. On the basis of the quenched case mentioned above the
deconfining transition should be order-disorder, and a genuine order
parameter should exist labeling the two phases. $N_c\to\infty$
arguments suggest indeed  that the symmetry involved as well as the
mechanism of confinement should be the same for quenched and unquenched
and $N_c$ independent. If for $N_f=2$ the chiral transition is second
order, then the transition for $m_q\ne0$ is a crossover, and the
deconfining transition is not order-disorder. A first order chiral
transition would instead be consistent with order-disorder, and possibly
be such up to
$m_q=\infty$ , which is the quenched case.

     The order of the transition can be investigated by looking at the
volume dependence of the specific heat in numerical simulations, by a
technique known as finite size scaling [see sect 2 below].
     The free energy density around the phase transition is
determined  by symmetry arguments \cite{LG} up to unknown numerical
coefficients, in terms of the order parameter. The (pseudo)critical
indices determined by looking at the susceptibility of the order
parameter must agree with the determination made by looking at the
specific heat . Such an agreement is needed to legitimate the choice of
the order paramenter.

     An order parameter $\langle\mu\rangle$ has been developed and tested
\cite{dg1}\cite{dg2}\cite{dg3}, based on the working hypothesis that the
mechanism of confinement is dual superconductivity of the
vacuum\cite{'tH2}.
$\langle\mu\rangle$ is the vacuum expectation value of an operator $\mu$
carrying magnetic charge, and , unless the Polyakov loop or the chiral
condensate is  well defined independent of
$N_c$,$N_f$ even in the continuum limit.

     The critical indices can be investigated by measuring susceptibilities
involving $\mu$ , in particular the quantity  $\rho=
d/d\tau[\ln(\langle\mu\rangle)]$ , with $\tau$ the reduced temperature
\begin{equation}
\tau = 1 - T/T_c
\end{equation}
If they agree with the determination made by use of the specific heat, an
additional
legitimation results for the order parameter and for the mechanism of
confinement by dual superconductivity. Confinement will then be defined
in terms of an appropriate symmetry, and the question whether the
deconfining transition and the chiral transitions occurr at the same
temperature becomes meaningful.

     In sect 2 we shall discuss new  results on the above
issues\cite{ddp},\cite{dddlpp} .

     In sect3 we  draw some conclusion.
\section{ Lattice results and Finite size scaling analysis.}
The theory of finite size scaling for higher order and weak first order
phase  transitions is based on renormalization group arguments
\cite{fisc}\cite{brez} ,which are expected to hold when, by approaching
the transition, the correlation lengths become much larger than the
lattice spacing. It allows to extrapolate to infinite volume results
obtained at finite volumes. The extrapolation depends on the
(pseudo)critical indices , which can then be determined from the volume
dependence, and with them the order and the universality class of the
transition.

    The relevant quantities , which are related to derivatives of
the free energy,and hence to the order of the transition, are the
susceptibilities . In the following we shall consider the susceptibility
of the chiral order parameter
\begin{equation}
     \chi = \frac{T}{V}\frac{\partial^2}{\partial m^2}\ln Z
\end{equation}
the specific heat
\begin{equation}
     c_V= \frac{1}{V T^2}\frac{\partial^2}{\partial(1/T)^2}\ln Z
\end{equation}
and
\begin{equation}
\rho=\frac{\partial}{\partial\tau}\ln\langle\mu\rangle
\end{equation}

All these susceptibilities will depend on the parameters which describe
the system, i.e.   $\beta=\frac{2N_c}{g^2}$  with  $g$  the gauge coupling,
and  $ma$  , the bare quark mass in units of the inverse lattice
spacing.

     The simulation is made on a lattice $N_t N_s^3$,$N_t\ll N_s$,  of extension
$N_t$ spacings along the time axis and $N_s$ along the three spatial
directions, with periodic boundary conditions (bc) in time for
bosons,antiperiodic bc for fermions.

The temperature is the inverse of
the time extension, or
\begin{equation}
              T = \frac{1}{N_t a(\beta,ma)}
\end{equation}
$a(\beta,ma)$ is the lattice spacing in physical units.

All the susceptibilities   as function of the
temperature will have peaks at $T_c$, which  diverge as  the volume
$V\to\infty$ with a power depending on the order of the transition
,specifically on the critical indices.  In the usual notation of
statistical mechanics,denoting by  L the spatial size of the system
\begin{eqnarray}
\chi &=&  L^{\gamma/\nu}\Phi_{\chi}(\tau L^{1\over\nu},mL^{y_h})\\
c_V &=&  c^0_V + L^{\alpha/\nu}\Phi_{c}(\tau L^{1\over\nu},mL^{y_h})\\
\rho &=& L^{1\over\nu}\Phi_{\mu}(\tau L^{1\over\nu},m L^{y_h})
\end{eqnarray}

     Since  $ T= {1\over N_ta(\beta,ma)} $
\begin{equation}
     \tau = 1 -
{a(\beta_c,0)\over a(\beta,ma)}
\end{equation}

     In the vicinity of the chiral point one can expand  in powers of
$\beta_c-\beta$ and of $m a$, obtaining

\begin {equation}
\tau \propto \beta_c - \beta  + K ma
\end{equation}
with
\begin{equation}
K = -
\left.\frac{\partial \ln a}{\partial (m a)}/\frac{\partial \ln 
a}{\partial \beta}
\right|_{\beta=\beta_c}
\end{equation}

In the quenched case the last term in eq(16) is absent.

When the volume of the system  goes large at $\tau L^{1\over\nu}$
fixed, the susceptibilities will tend to a finite limit certainly if the
transition is second order, and the free energy is continuous and finite
in the neighbourhood of the chiral point.
     This implies that the powers of L in front of the susceptibility eq's
(12),(13),(14) have to be eliminated by the dependence on  $m L^{y_h}$,
and this means for the three susceptibilities
\begin{eqnarray}
\chi&\propto& m^{-{\gamma\over{\nu y_h}}} F_{\chi}(\tau L^{1\over\nu})\\
c_V-c^0_V &\propto&
    m^{-{\alpha\over{\nu y_h}}} F_c(\tau L^{1\over\nu})\\
\rho &\propto& m^{-{1\over{\nu y_h}}} F_{\rho}(\tau L^{1\over\nu})
\end{eqnarray}

     If the transition is
weak first order the system will behave in the same way
    for intermediate values of L,
smaller or equal than the critical  correlation length, apart for the
different value of the critical indices.

Eq's(18,(19),(20)  imply that the maxima
of the susceptibilities with respect to
$\tau$ lie on the line of the plane  $(\beta,ma) $ of equation
\begin{equation}
\tau L^{1\over\nu} = C
\end{equation}
with $C$ a constant,
or, by use of eq(16)
\begin{equation}
\beta_c - \beta  + K ma - {C\over L^{1\over\nu}} =0
\end{equation}

This relation can be tested against the lattice data , to determine the
value of $\nu$ . We will discuss the result in the following .

     Eqs (18),(19),(20)  also imply that height of the the maximum of the
susceptibilities is volume independent, whatever the order of the
transition, in the limits of the regime in which these equations are
obeyed when the transition is first order.

     For second order transition the result is valid for any volume and fits
     the idea that the transition line in the plane $(\beta,ma)$ is a
crossover.

     For a first order transition ,if we cross the transition line at fixed
$ma$, we expect a behaviour typical of a first order transition, namely
a growth of the peaks of the susceptibilities proportional to the volume.
Indeed for a first order transition $\alpha=\gamma=1$,$\nu=1/d =1/3$.
     This behaviour will be visible when L  becomes larger than the critical
correlation length, and together with it the typical two-peak
distribution of the internal energy will appear.

     Moreover, if we keep  $mL^{y_h}$ constant, eqs(12),(13),(14)) require
that the maximum of the peaks is again proportional to a well defined
power of  L  depending on the nature of the transition.

     The details of the simulations will be reported elsewhere
\cite{ddp}\cite{dddlpp} Here we only summarize the results.

    As a strategy
we assume the critical indices expected for second order O(4) phase
transition and in alternative those of a first order transition and we
investigate wether data are consistent with either of them. A measure of
the agreement is the value of $\chi^2/dof$.

In table 1 we recall values of
the critical indices for the two cases.

We  investigate
\begin{itemize}
\item[1)] The position of the peak as a function of mass and volume , which
scales according to Eq(22). The  $\chi^2/dof$  is $\sim 1$  for
the choice first order,  and  typically $\sim 10$    for O(4) second order.
\item[2)] The peak values of  the susceptibilities. They all occurr on the same
line, within errors.The height of the peak is volume independent at fixed
$ma$ for moderate values of the volume and proportional to an
appropriate power of $(ma)$ Eqs(18)-(20), which depends on the nature of
the transition. Here again the   $\chi^2/dof$ is compatible with 1 for a
first order transition , and much worse , typically [
$\chi^2/dof\approx10 $] for O(4) second order.
\item[3)] For low values of m (e.g. $ma =
.0135$ ) as the volume increases further a growth of the height of the
peak is observed and some sign of bistability in the time histories of the
energy density. An increase of the heigth of the peak by a factor
$\approx2$ is observed going from $16^3$ to $32^3$ . This means that there
is a transition, and not a crossover. At higher values of $ma$ the
transition looks weaker and we were not yet able to reach a  large enough
volume to see a bistability .
\item[4)] The behaviour with respect to
volume at fixed
$mL^{y_h}$ is consistent with first order transition , and disfavours
second order : typically $\chi^2/dof\approx 1 $  for first order,
$\chi^2/dof\approx10 $  for second order O(4).
\end{itemize}
Previous investigations of the same system were made on
rather small  volumes\cite{ts1}\cite{kar3} \cite{MILCH} or in a less
systematic way,and the results where admittedly inconclusive, although a
slight psicological preference appeared for a second order transition .
Numerically our determinations are consistent with the previous ones when
done at the same values of the parameters. The details of the simulations
and the comparison with previous work will be presented elsewhere
\cite{ddp}\cite{dddlpp}.

\section{ Discussion.}
We have obtained substantial preliminary evidence that the chiral
transition for
$N_f=2$ QCD is first order.
     This makes the deconfining transition consistent with an order disorder
transition.
     The disorder parameter $\langle\mu\rangle$ detecting dual
superconductivity of the vacuum provides a determination of the critical
indices consistent with the one based on the specific heat,and  a
transition line consistent with the other susceptibilities. It can then
be used as an order parameter for confinement.
     The deconfining transition and the chiral transition occurr at the same
T.
     A careful numerical study of the U(1) axial anomaly across the
deconfining transition is on the way and will provide a check of the
assumption made in \cite{PW} that the relevant degrees of freedom at the
chiral transition are the pseudoscalar goldstone particles.

I wish to thank all my collaborators ,  M. D'Elia, B.Lucini,
G. Paffuti, C. Pica who contributed substantially to the results
presented. Many ideas and motivations also come from discussions and
collaboration with Yu. A. Simonov , to whom this paper is dedicated with
deep friendship in the occasion of his 70th birthday.

\newpage
%
\begin{table}
\caption{}{The (pseudo) critical
indices for  weak  1st order and second
order $O(4)$ phase transition.
}

\bigskip
\begin{tabular}{|c|c|c|c|c|}\hline
  &  $\alpha $ &   $  \gamma   $ & $   \nu  $ &  $y_h$ \\ \hline
1st order & 1 & 1 & 1/3 &3\\
$O(4)$ 2nd order &  -.25(1)   &      1.47(1)  &     .75(1) & 
2.49(1) \\ \hline
\end{tabular}
\end{table}

\end{document}